# Methods of Electron Microscopy of Biological and Abiogenic Structures in Artificial Gas Atmospheres

O. V. Gradov[a, b] and M. A. Gradova[b]

[a]Tal'roze Institute of Energy Problems of Chemical Physics, Russian Academy of Sciences,
Leninskii pr. 38, Moscow, 119334 Russia

[b]Semenov Institute of Chemical Physics, Russian Academy of Sciences, ul. Kosygina 4, Moscow, 119991 Russia
e-mail: o.v.gradov@gmail.com

**Abstract**—This paper reviews opportunities for using electron microscopy in various gas atmospheres for the analysis and morpho-physiological modification of biological structures. The approaches that allow varying the gaseous phase content, as well as temperature, humidity, and pressure, are considered. The applicability of both kinetic and dynamic approaches to the tissue and bioinorganic structure manipulations is pointed out. The possibility of simulation of the beam-induced formation and disintegration of abiogenetic molecular structures is also mentioned as a particular case of the electron beam influence and treatment of the precursor medium in an artificial atmosphere.

## 1. PRINCIPLES OF ELECTRON MICROSCOPY IN A GAS ATMOSPHERE

In the last quarter of the 20th century, the ultimate problem and exceptional necessity for experimental morphological studies in situ under an electron beam affecting singles cells [1] and tissues [2] was to create a gas atmosphere in a chamber suitable for the intravital observation and retention of physiological functions of the object under study, which conflicted with the accessibility of the quality of detection of the image or analytical response from the experimental tissue because of the inevitableness of electron diffusion in the gas [3], from which followed not only the interaction of the irradiating beam with the atmosphere surrounding the biological object but also the interaction of the signal beam, i.e., the beam being detected, with the gas [4], which made senseless any attempts to acquire analytical information from the registered signal as a result of the effects of electron–ion recombination [5] and background problems with cathodoluminescence and scintillations in the gas [6]. From these prerequisites, there was the critical requirement in practice to overcome the conflict between the intravital character of the physiological study of the object in a natural environment that to this end was to be had in the electron microscope chamber (the composition of the gas, variable temperature and pressure, possibility for the partial replacement of the composition of the gas for the purposes of preconditioning [7–9], among other things, in the case of the change in the pressure in the chamber and partial pressure of the gas [10, 11], etc.) and needs of registration, which were integral features of the method itself for physical reasons (or, more properly, principles).

In addition, in the absence of the possibilities for microscopy in various gas atmospheres, there was also no possibility for the intravital dynamic study of the structure of cells and tissues depending on the change in the composition of the experimental "atmosphere" (which provided the basis for such a currently promising field as gas biology [12, 13]) despite the fact that methods for dynamic and, in particular, intravital structural study have been known in electron microscopy since the mid-1960s (so-called stroboscopic electron microscopy [14–17], which is currently successfully replaced by the methods of four-dimensional electron microscopic analysis [18–20], although many such methods of microscopy, in particular, dynamic cryoelectron microscopy and microprocessing, which is a novel approach in structural, systemic, and synthetic biology [21], retain the elements of a classic stroboscopic technique [22]).

The solution was reached in the pioneer works by G.D. Danilatos, who laid down the principles of a new method, so-called ESEM (environmental scanning electron microscopy), which allowed introducing samples into a controlled gas atmosphere, including natural atmosphere. Omitting the intermediate phases of the publication activities of the aforesaid author, who has over fifty works on ESEM, one can cite a series of works describing a generalized form of equipment of this type, which is suitable for both atmospheric scanning microscopy and similar electron microscopy in a controlled gas atmosphere, for the sake of completeness of information [23–26]. A little later, similar units became quite common in various laboratories, which resulted in the creation of many techniques for fine ultrastructural analysis or cytomorphological analysis under atmospheric pressure. Today, there are many methods and techniques of this type, including atmospheric pressure scanning transmission electron microscopy [27], atmospheric correlative scanning electron microscopy and atmospheric correlative immunoelectron microscopy [28, 29], and technologies for atmospheric scanning electron microscopy for the registration of dynamic phenomena in liquids and gases on the corresponding interphase boundaries [30]. These technologies are suitable for the analysis of single cells in a gas atmosphere [31, 32], cell cultures and tissues [33, 34], and visualization and systematization of biologically relevant objects of the natural environment [35] and protein microcrystals [36] (lately, effective visualization down to the atomic level has been achieved under normal pressure of the natural environment [37]).

## 2. BIOPHYSICAL AND BIOCHEMICAL PROBLEMS OF ELECTRON MICROSCOPY IN A GAS ATMOSPHERE

Two fundamental elements of atmospheric electron microscopy and electron microscopy in a gas atmosphere for biological applications can be distinguished: (a) achieving variable pressure of the atmosphere being pumped and (b) selecting a biologically optimum chemical composition of the atmosphere. The first attempts at creating electron microscopes with controlled pressure of the gas date back to the early 1960s [38], but their actualization did not take place until the creation of ESEM in 1980s, while currently variable pressure electron microscopes are heavily used for charge contrast mapping of biological samples [39] and similar tasks of biological visualization [40], as well as for ultramicroscopic sample preparation and processing of samples at a nanostructured level [41]. The second problem is somewhat more complicated because, if electron microscopy is considered in a reactor approximation [42], it is necessary to take into account the inevitable interaction of the gas being pumped into the chamber with the gaseous components of biological tissues, which is particularly relevant in the case of objects popular for electron microscopic visualization such as gas vacuoles of single-celled animals [43] or lung tissues [44], as well as in experiments on the study of the secondary structure of protein in gas vesicles [45] and applied test measurements on gas permeable lenses [46]. Since numerous myohistological studies and interdisciplinary biomolecular works are also conducted in electron micro-scope chambers with a controlled composition of the gas [47, 48], this statement can be extended to this field as well, and taking into account the catalytic character of interactions and enzymatic properties of myosin [49], this line of research can be interpreted as an analog of catalytic electron microscopic studies in a gas atmosphere [50]. It is clear that, since catalytic processes in a biological medium occur under the principles of biophysical chemistry [51] and enzyme kinetics [52], the transition to studying the microstructure and ultrastructure in naturally occurring atmospheric conditions that preserve the dynamics is equivalent to the emergence of electron microscopic dynamic biochemistry, not any longer in that quite simplified interpretation that existed in the 1940s–1950s [53, 54] and affected the scientific thought up to the late 1980s [55], but in that more modern sense that is associated with chemical and macromolecular (even phase supramolecular) analyses [56], compartmentalization, and chemical kinetics, in particular, modern enzyme kinetics [57].

Dynamic physicochemical processes can be studied under the thermostatting of the experimental atmosphere at a specified level. Biophysical and physiological processes also require thermostatting (if we are not referring to cryobiology and extremophiles). Accordingly, electron microscopic accessories and tools for the preparation and management of a sample in situ in a chamber with thermostatting/cryostatting are required.

Differentiated tools for the monitoring of the atmosphere in a gas chamber are currently developed: gas injection holders with thermal heating in transmission microscopy [58], special-purpose cells for in situ observations [59, 60], multifunctional scanning electron microscope systems for work in a controlled atmosphere [61], which progressively differ from earlier constructions [62] (in particular, by the wide-range temperature control with the retention of the content of gases in an analytical chamber [63]). Multiple metrological procedures applicable also in the case of work with biological or hybrid objects have been developed specifically for this equipment. This refers to the methods of semiquantitative X-ray microanalysis with digital data processing which evens out the aberrations introduced by the gas [64], specialized atmospheric technologies that correct aberrations [65], techniques for measuring the effective path length of charge carriers of a scanning beam in a gas atmosphere [66], gas cascade amplification in high-resolution electron microcopy [67], and methods for real-time control of the composition of a gas in a chamber during low-energy electron spectroscopy or chemical visualization performed on the basis of equivalent principles [68]. The highest accuracy of metrology is achieved in the position-sensitive analysis of such a type when using ideal experimental units with electron emission from carbon tubes excited under gaseous and atmospheric conditions by atmospheric-pressure plasma [69].

## 3. PROBLEMS OF HUMIDITY OF BIOLOGICAL SAMPLES AND OXIDATIVE NATURE OF THE ATMOSPHERE

In all the biological works on atmospheric-pressure electron microscopy, there is a common problem, humidity of samples (similar to how the quality of the surface depends on the wettability of a crystal and can be qualitatively determined by it in X-ray structural and X-ray diffraction analyses [70]). As is apparent from psychrometric principles [71], the rate of evaporation of a liquid increases with the decrease in the pressure and relative humidity; hence, it is necessary to use a chamber that maintains a certain humidity, but this may conflict with the parameters of a gas column and/or chamber which are optimum for the registration. At the same time, the need for the psychrometric optimization for cells, tissues, or isolated fragments can be considered to be proven (at least, long and thoroughly investigated for plants [72, 73] and postulated on the basis of obvious considerations for the climatic stabilization during the cultivation of animal tissues in thermostatted incubators with a controlled gas atmosphere [74–76]).

Since the early 1970s [77, 78], works on the constructions of atmospheric chambers for the analysis of wet biological samples have appeared. Currently, this trend continues to develop but already with a focus on nanostructured materials science [79]. Recent advances in the field of microstructural aquametry are associated with cryoelectron microscopy [80], especially with the correlative analysis of induced changes under nucleolar stress [81]. When natural atmosphere or artificial gas filling of chambers is present, aquametry differs from known thermal vacuum methods of aquametry [82] because, from the psychrometric perspective, it possesses higher complexity and dimension of curves depending on the content of gases. The correlation of this characteristic with the wettability of the surface and its sorption characteristics is an interesting problem because hydrophobicity affects the formation of the condensate and the results of the electron microscopic measurement of the dew point in the absence of other parameters [83].

Sorption and filtration specimens and chromatographic stationary phases obtained as a result of biochemical and biomolecular research are some of the particular cases of such difficulties. In particular, changes in the structure of sorbents during gas chromatography [84, 85] and changes in analytical plates, which are based on reactions with a solid surface, are known [86]. Moreover, there are specialized electron microscope units for obtaining isotherms of adsorption [87] and data on structural changes at an atomic level during the interaction of a gas with a solid surface [88] and for observing processes of interaction in real time [89, 90]. Given that most cytomorphological and ultrastructural studies are generally conducted on supports and electron microscopy grids, the need for the registration of aforementioned phenomena in atmospheric electron microscopy with the use of wet specimens is obvious.

Another biological fine point for conducting electron microscopy in gas atmospheres is the oxidative character of many atmospheres, including those belonging to standard atmospheres for gas pumping. Thus, effects of gas polymerization during electron or other charge-corpuscular bombardment of surfaces resulting in morphologically detectable changes have been known since the 1940s [91], while technologies for the ablation of biological surfaces using activated oxygen were developed for electron microscopy in the 1970s [92]. In addition, despite a clear focus on nanostructured materials science, the range of problems of synthesis or processing of matter (cells, tissues, biological materials, etc.) with a focused electron or ion beam [93, 94], which results in the preparation of many exotic intermediates and assembly of exotic structures, continues to remain relevant. Multiple attempts to minimize and reduce interactions of the beam with the gas in controlled-pressure electron microscopes [95] ultimately lead to the reduction in chemical processes in the gas atmosphere and on the surface but do not exclude certain induced structural transformations in the substance itself.

Physical optimization using Monte Carlo methods, which is often used by experts in electron microscopy when simulating electron beam dispersion [96], where the sample acts as a target on the propagation path of electrons, with which exactly the beam interacts, does not provide an answer to the question on the structural–chemical transformations in the atmosphere if they are not associated with the elementary chemism (isotopy and nuclear physical processes). Generally, such simulations affect the improvement of the quality of the image [97] but not the biologically orientated optimization of the physics of the process of interaction of the beam with the tissue, although it is known in biomedicine that the rational configuration of an electron beam provides the optimization of the structural–biological effect [98], while methods for site-specific beam focusing have long been used in cryoelectron microscopy [99]. A so-called surface skirt, beam broadening during its dispersion in a gas atmosphere [100], can be a partial solution in the method of electron microscopy of biological specimens, although the decrease in focusing causes the deterioration of the quality of the image, and the latter is fraught with a well known effect on the results of semiquantitative X-ray microanalysis in a column [101]. In turn, problems with distinguishing the components of the signal in scanning electron microscopy under the conditions of an atmosphere (a gas) or natural atmosphere and at "natural" temperature have been known since the 1970s [102], while methods for their distinguishing emerged only in the early 1990s—no sooner than the tools of computer automation made it possible to do so [103].

## 4. APPLICABILITY OF GAS ELECTRON MICROSCOPY IN THE STUDY OF ABIOGENESIS

From the viewpoint of the corpuscular effect on biological and abiogenic media [104, 105], a synchronized synthesis and chemical analysis of protobiopolymers under a beam directly in a gas column, which simulates the conditions for abiogenesis and formation of prebiological compounds under the action of known corpuscular factors in various gas atmospheres, can be of interest. In this case, real-time analysis can be conducted according to the principles of position-sensitive electron microprobe analysis/characteristic X-ray secondary-emission spectrometry (energy-dispersive or wavelength-dispersive spectrometry depending on whether photons are scaled by energy or by wavelength). Also, despite the fact that most protobiopolymers formed in thin layers or films are "soft matter," micro X-ray structural analysis of formed entities is implementable [106]. From the chemical perspective, the task of simulating abiogenetic processes in artificial gas atmospheres is determined by the fact that known or proposed compositions of prebiological and xenobiotic atmospheres, in particular, reducing methane-containing and hydrogen-containing atmospheres [107], volcanic eruptive and fumarole gases [108], nitrogen-containing atmosphere in a series of exotic exoplanetary concepts [109], and a whole range of others, contain various combinations of simple organogen compounds such as ammonia, methane, hydrogen, nitrogen, carbon monoxide, carbon dioxide, sulfur dioxide, and formaldehyde required for the abiogenic synthesis of organic matter. In a general thermodynamic case, an abiogenic atmosphere is an ideal gas with which a protocell membrane or lipid bilayer can "work" as Maxwell's demon [110]. From the perspective of physics of abiogenesis, this problem statement is determined by a known role of cosmic rays [104, 105], in particular, charged particles [111, 112], in the formation of prebiological organic compounds in gas atmospheres [113], which is fundamentally different from the principles of cosmic formation of the same compounds under the action of cosmic radiations [114]. Much the same is true for the emission of particles by radioisotopes [115], which, taking into account the presence of the phenomena of cluster radioactivity, emission of one or two protons, positron decay, etc., makes it possible to bombard a "precursor" for abiogenic synthesis with modes of radiations which are quite differentiated for the synthesis of various products. Also, because of the possible catalytic role of the mineral support in the processes of prebiological synthesis and structure formation, it is essential to take into account the chemical interaction of the atmosphere with the support when studying such processes in an electron microscope chamber. However, this aspect of the problem is not included in the goals of this paper; therefore, it is not considered further in more detail. As a result of the reductive character of the primitive atmosphere (which corresponds to the necessary conditions for the abiogenic synthesis of organic matter), there should be no problems related to the fixation of samples in a gas atmosphere associated with gas oxidation (see i. 3) during the simulation of abiosynthesis in a gas atmosphere in an electron microscope column.

### NECESSARY TERMINOLOGICAL REMARK

In conclusion, it should be noted that, while not being a tool for nondestructive control, a method of gas electron microscopy operates as an analytical method only to the limits determined by the quantum structure of possible precursors and protobionts [116, 117], after which it starts to directly affect the sample itself, thus causing organization or transformations in its structure. That said, we would also like to warn readers of careless citation without using an original work, which is common for periodical publications in Russian, when microscopes capable of registering the behavior of a quantum gas [118, 119] are "intermixed" with gas electron microscopes similar to the microscopes considered in this work, as a result of which statements are made about the achievability of quantum resolution on gas electron microscopes, which is not implementable in practice because of the interaction with the atmosphere in the propagation path of the beam to the object or else, it is more correct to say, when referring to the action by the beam on the precursor specimen, the target.

### CONCLUSIONS

(1) The problem of environmental condition microscopy can be divided into two parts, namely, microscopy at natural (room/laboratory) temperature and microscopy in a natural atmosphere. Technically, both problems can be solved in one instrument combined for these purposes. The latter problem can also be extrapolated to other atmospheres that are not equivalent to Earth's atmosphere with respect to the composition. Under such conditions, abiogenesis can be directly simulated under a beam of an electron microscope, or xenobiological systems, chemoautotrophic bacteria (including anaerobic bacteria), lability of microbic ecosystems for atmospheres with various compositions of the gas, etc., can be studied. The possibility for the occurrence of compartmentalized structures from abiogenic material as early as the early stages of formation of protobiologic systems under the corresponding conditions of the primitive atmosphere of Earth makes the morphostructural study of the products of abiogenic synthesis a necessary addition to the data on their chemical composition.

(2) The problem of deep environmental electron microscopy in liquid media has not been solved and, probably, is unsolvable at this stage (although, e.g., on accelerators, sources of synchrotron radiation, corresponding leads are possible; however, these are not electron leads but leads of the "optical" beam).

(3) It is possible to combine morphological ("localomics"), dynamic ("dynamomics"), spectral (X-ray secondary-emission spectrometry or microprobe in a scanning mode, which belongs to localomics with respect to the time scale), and structural (electron diffraction and X-ray diffraction methods in situ; in general, structural analysis in situ of soft matter structures such as biological systems) analyses in the case of work and manipulations in a gas atmosphere.

(4) The combination of localomics and dynamomics results in synchronic morpho-physiological/functional-morphological and morpho-biochemical/histochemical analyses in situ under specified conditions of the atmosphere, which creates prerequisites for conceptually new 4D ultramorphology on the basis of gas electron microcopy.

(5) It is possible to combine synthesis or induced self-assembly of organic structures and compounds under a beam in an atmosphere, which makes it possible to apply the method of gas electron microscopy for simulating chemical stages of abiogenesis and other beam-induced processes using a gas-filled chamber as an experimental reactor.

## ACKNOWLEDGMENTS

We are grateful to the members of the Department of Metrology and Measuring Instruments, Vernadsky Institute of Geochemistry and Analytical Chemistry, Russian Academy of Sciences, who provided us with the documentation for unique experimental equipment that offered an opportunity to start works in this direction.

## REFERENCES

1. Miller, J.H., Sowa Resat, M., Metting, N.F., Wei, K., Lynch, D.J., and Wilson, W.E., Monte Carlo simulation of single-cell irradiation by an electron microbeam, *Radiat. Environ. Biophys.,* 2000, vol. 39, no. 3, pp. 173−177.

2. Miller, J.H., Suleiman, A., Chrisler, W.B., and Sowa, M.B., Simulation of electron-beam irradiation of skin tissue model, *Radiat. Res.,* 2011, vol. 175, no. 1, pp. 113−118.

3. Moncrieff, D.A., Barker, P.R., and Robinson, V.N.E., Electron scattering by gas in the scanning electron microscope, *J. Phys. D: Appl. Phys.,* 1979, vol. 12, no. 4, pp. 481−488.

4. Mathieu, C., The beam-gas and signal-gas interactions in the variable pressure scanning electron microscope, *Scanning Microsc.,* 1999, vol. 13, no. 1, pp. 23−41.

5. Morgan, S.W. and Phillips, M.R., Transient analysis of gaseous electron-ion recombination in the environmental scanning electron microscope, *J. Microsc.* (Oxford, U. K.), 2006, vol. 221, no. 3, pp. 183−202.

6. Danilatos, G.D., Cathodoluminescence and gaseous scintillation in the environmental SEM, *Scanning,* 1986, vol. 8, no. 6, pp. 279−284.

7. Liu, W., Liu, Y., Chen, H., Liu, K., Tao, H., and Sun, X., Xenon preconditioning: molecular mechanisms and biological effects, *Med. Gas Res.,* 2013, vol. 3(1), no. 3. doi: 10.1186/2045-9912-3-3

8. Delgado-Roche, L., Martínez-Sánchez, G., and Re, L., Ozone oxidative preconditioning prevents atherosclerosis development in New Zealand white rabbits, *J. Cardiovasc. Pharmacol.,* 2013, vol. 61, no. 2, pp. 160−165.

9. Smit, K.F., Oei, G.T., Brevoord, D., Stroes, E.S., Nieuwland, R., Schlack, W.S., Hollmann, M.W., Weber, N.C., and Preckel, B., Helium induces preconditioning in human endothelium in vivo, *Anesthesiology,* 2013, vol. 118, no. 1, pp. 95−104.

10. Duan, Z., Zhang, L., Liu, J., Xiang, X., and Lin, H., Early protective effect of total hypoxic preconditioning on rats against systemic injury from hemorrhagic shock and resuscitation, *J. Surg. Res.,* 2012, vol. 178, no. 2, pp. 842−850.

11. Soejima, Y., Ostrowski, R.P., Manaenko, A., Fujii, M., Tang, J., and Zhang, J.H., Hyperbaric oxygen preconditioning attenuates hyperglycemia enhanced hemorrhagic transformation after transient MCAO in rats, *Med. Gas Res.,* 2012, vol. 2(1), no. 9. doi: 10.1186/2045-9912-2-9

12. Semenza, G.L. and Prabhakar, N.R., Gas biology: Small molecular medicine, *J. Mol. Med.* (Heidelberg, Ger.), 2012, vol. 90, no. 3, pp. 213−215.

13. Nakao, A. and Toyoda, Y., Book review: Gas biology research in clinical practice, edited by Toshikazu Yoshikawa and Yuji Naito, *Med. Gas Res.,* 2011, vol. 1(1), no. 25. doi: 10.1186/2045-9912-1-25

14. Frey, S.A., Stroboscopic technique for electron micrographs, *Med. Biol.,* 1965, vol. 15 (Suppl.), pp. 19−21.

15. Plows, G.S. and Nixon, W.C., Stroboscopic scanning electron microscopy, *J. Sci. Instrum.,* 1968, vol. 2, no. 1, pp. 595−600.

16. Robinson, G.Y., Stroboscopic scanning electron microscopy at gigahertz frequencies, *Rev. Sci. Instrum.,* 1971, vol. 42, no. 2, pp. 251−255.

17. Szentesi, O.I., Stroboscopic electron mirror microscopy at frequencies up to 100 MHz, *J. Phys. E: Sci. Instrum.,* 1972, vol. 5, no. 6, pp. 563−567.

18. Zewail, A.H., Four-dimensional electron microscopy, *Science,* 2010, vol. 328, no. 5975, pp. 187−193.

19. Kwon, O.-H., Ortalan, V., and Zewail, A.H., Macromolecular structural dynamics visualized by pulsed dose control in 4D electron microscopy, *Proc. Natl. Acad. Sci. U. S. A.,* 2011, vol. 108, no. 15, pp. 6026−6031.

20. Baskin, J.S., Park, H.S., and Zewail, A.H., Nanomusical systems visualized and controlled in 4D electron

microscopy, *Nano Lett.,* 2011, vol. 11, no. 15, pp. 2183−2191.

21. Gradov, O.V. and Gradova, M.A., Cryo-electron microscopy as a functional instrument for systems biology, structural analysis and experimental manipulations with living cells. A comprehensive analytical review of the current works, *Probl. Cryobiol. Cryomed.,* 2014, vol. 24, no. 3, pp. 193−211.

22. Nejadasl, K.F., Karuppasamy, M., Koster, A.J., and Ravelli, R.B., Defocus estimation from stroboscopic cryo-electron microscopy data, *Ultramicroscopy,* 2011, vol. 111, no. 11, pp. 1592−1598.

23. Danilatos, G.D., Design and construction of an atmospheric or environmental SEM (part 1), *Scanning,* 1981, vol. 4, no. 1, pp. 9−20.

24. Danilatos, G.D. and Postle, R., Design and construction of an atmospheric or environmental SEM (part 2), *Micron,* 1983, vol. 14, no. 1, pp. 41−52.

25. Danilatos, G.D., Design and construction of an atmospheric or environmental SEM (part 3), *Scanning,* 1985, vol. **7**, no. 1, pp. 26−42.

26. Danilatos, G.D., Design and construction of an atmospheric or environmental SEM (part 4), *Scanning,* 1990, vol. 12, no. 1, pp. 23−27.

27. De Jonge, N., Bigelow, W.C., and Veith, G.M., Atmospheric pressure scanning transmission electron microscopy, *Nano Lett.,* 2010, vol. 10, no. 3, pp. 1028−1031.

28. Morrison, I.E.G., Dennison, C.L., Nishiyama, H., Suga, M., Sato, C., Yarwood, A., and O'Toole, P.J., Atmospheric scanning electron microscope for correlative microscopy, *Methods Cell Biol.,* 2012, vol. 111, ch. 16, pp. 307−324.

29. Maruyama, Y., Ebihara, T., Nishiyama, H., Suga, M., and Sato, C., Immuno EM-OM correlative microscopy in solution by atmospheric scanning electron microscopy (ASEM), *J. Struct. Biol.,* 2012, vol. 180, no. 2, pp. 259−270.

30. Suga, M., Nishiyama, H., Konyuba, Y., Iwamatsu, S., Watanabe, Y., Yoshiura, C., Ueda, T., and Sato, C., The atmospheric scanning electron microscope with open sample space observes dynamic phenomena in liquid or gas, *Ultramicroscopy,* 2011, vol. 111, no. 12, pp. 1650−1658.

31. Sato, C., Manaka, S., Nakane, D., Nishiyama, H., Suga, M., Nishizaka, T., Miyata, M., and Maruyama, Y., Rapid imaging of mycoplasma in solution using atmospheric scanning electron microscopy (ASEM), *Biochem. Biophys. Res. Commun.,* 2012, vol. 417, no. 4, pp. 1213−1218.

32. Murai, T., Sato, M., Nishiyama, H., Suga, M., and Sato, C., Ultrastructural analysis of nanogold-labeled cell surface microvilli in liquid by atmospheric scanning electron microscopy and their relevance in cell adhesion, *Int. J. Mol. Sci.,* 2013, vol. 14, no. 10, pp. 20809−20819.

33. Nishiyama, H., Suga, M., Ogura, T., Maruyama, Y., Koizumi, M., Mio, K., Kitamura, S., and Sato, C., Atmospheric scanning electron microscope observes cells and tissues in open medium through silicon nitride film, *J. Struct. Biol.,* 2010, vol. 169, no. 3, pp. 438−449.

34. Nishiyama, H., Suga, M., Ogura, T., Maruyama, Y., Koizumi, M., Mio, K., Kitamura, S., and Sato, C., Reprint of: Atmospheric scanning electron microscope observes cells and tissues in open medium through silicon nitride film, *J. Struct. Biol.,* 2010, vol. 172, no. 2, pp. 191−202.

35. Luo, P., Morrison, I., Dudkiewicz, A., Tiede, K., Boyes, E., O'Toole, P., Park, S., and Boxall, A.B., Visualization and characterization of engineered nanoparticles in complex environmental and food matrices using atmospheric scanning electron microscopy, *J. Microsc.* (Oxford, U. K.), 2013, vol. 250, no. 1, pp. 32−41.

36. Maruyama, Y., Ebihara, T., Nishiyama, H., Konyuba, Y., Senda, M., Numaga-Tomita, T., Senda, T., Suga, M., and Sato, C., Direct observation of protein microcrystals in crystallization buffer by atmospheric scanning electron microscopy, *Int. J. Mol. Sci.,* 2012, vol. 13, no. 8, pp. 10553−10567.

37. Creemer, J.F., Helveg, S., Hoveling, G.H., Ullmann, S., Molenbroek, A.M., Sarro, P.M., and Zandbergen, H.W., Atomic-scale electron microscopy at ambient pressure, *Ultramicroscopy,* 2008, vol. 108, no. 9, pp. 993−998.

38. Heide, H.G., Electron microscopic observation of specimens under controlled gas pressure, *J. Cell Biol.,* 1962, vol. 13, no. 1, pp. 147−152.

39. Clode, P.L., Charge contrast imaging of biomaterials in a variable pressure scanning electron microscope, *J. Struct. Biol.,* 2006, vol. 155, no. 3, pp. 505−511.

40. Griffin, B.J., Variable pressure and environmental scanning electron microscopy: Imaging of biological samples, *Electron Microscopy Methods and Protocols,* Methods Mol. Biol., Totowa, NJ: Humana Press, 2007, vol. 369, 2 ed., pp. 467−495.

41. Niitsuma, J.-I., Sekiguchi, T., Yuan, X.-L., and Awano, Y., Electron beam nanoprocessing of a carbon nanotube film using a variable pressure scanning electron microscope, *J. Nanosci. Nanotechnol.,* 2007, vol. **7**, no. 7, pp. 2356−2360.

42. Kliewer, C.E., Kiss, G., and DeMartin, G.J., Ex situ transmission electron microscopy: A fixed-bed reactor approach, *Microsc. Microanal.,* 2006, vol. 12, no. 2, pp. 135−144.

43. Houwink, A.L., Flagella, gas vacuoles and cell-wall structure in *Halobacterium halobium;* an electron microscope study, *J. Gen. Microbiol.,* 1956, vol. 15, no. 1, pp. 146−150.

44. Maina, J.N., Scanning electron microscope study of the spatial organization of the air and blood conducting components of the avian lung (*Gallus gallus* variant domesticus), *Anat. Rec.,* 1988, vol. 222, no. 2, pp. 145−153.

45. McMaster, T.J., Miles, M.J., and Walsby, A.E., Direct observation of protein secondary structure in gas vesicles by atomic force microscopy, *Biophys. J.,* 1996, vol. 70, no. 5, pp. 2432−2436.

46. Fowler, S.A., Korb, D.R., Finnemore, V.M., Ross, R.N., and Allansmith, M.R., Coatings on the surface of siloxane gas permeable lenses worn by keratoconic patients: a scanning electron microscope study, *CLAO J.,* 1987, vol. 13, no. 4, pp. 207−210.

47. Sugi, H., Akimoto, T., Chaen, S., and Suzuki, S., ATP-induced axial movement of myosin heads in living thick filaments recorded with a gas environmental chamber attached to the electron microscope, *Adv. Exp. Med. Biol.*, 1998, vol. 453, pp. 53–62.

48. Minoda, H., Okabe, T., Inayoshi, Y., Miyakawa, T., Miyauchi, Y., Tanokura, M., Katayama, E., Wakabayashi, T., Akimoto, T., and Sugi, H., Electron microscopic evidence for the myosin head lever arm mechanism in hydrated myosin filaments using the gas environmental chamber, *Biochem. Biophys. Res. Commun.*, 2011, vol. 405, no. 4, pp. 651–656.

49. Engelhardt, W.A. and Ljubimowa, M.N., Myosine and adenosinetriphosphatase, *Nature*, 1939, vol. 144, no. 3650, pp. 668–669.

50. Gai, P.L., Environmental high resolution electron microscopy of gas-catalyst reactions, *Top. Catal.*, 1999, vol. 8, nos. 1–2, pp. 97–113.

51. Swiegers, G., *Mechanical Catalysis: Methods of Enzymatic, Homogeneous, and Heterogeneous Catalysis*, Hoboken, NJ: Wiley, 2008.

52. Steinfeld, J.I., Francisco, J.S., and Hase, W.L., *Chemical Kinetics and Dynamics*, Upper Saddle River, NJ: Prentice Hall, 1998.

53. White, A., Dynamic aspects of biochemistry, *Yale J. Biol. Med.*, 1947, vol. 19, no. 5, p. 900.

54. Cammarata, P.S., Dynamic aspects of biochemistry, *Yale J. Biol. Med.*, 1953, vol. 25, no. 6, pp. 546–547.

55. Turning over an old leaf: *Dynamic Aspects of Biochemistry* by Ernest Baldwin. pp. 457. Cambridge University Press, 1947, *Biochem. Educ.*, 1988, vol. 16, no. 3, p. 180.

56. Benesch, J.L.P. and Ruotolo, B.T., Mass spectrometry: Come of age for structural and dynamical biology, *Curr. Opin. Struct. Biol.*, 2011, vol. 21, no. 5, pp. 641–649.

57. Gradova, N.B., A review of the textbook *Osnovy dinamicheskoi biokhimii* (Fundamentals of Dynamic Biochemistry) by V.K. Plakunov and Yu.A. Nikolaev (Moscow: Logos, 2010), *Microbiology*, 2011, vol. 80, no. 2, p. 273.

58. Kamino, T., Yaguchi, T., Konno, M., Watabe, A., Marukawa, T., Mima, T., Kuroda, K., Saka, H., Arai, S., Makino, H., Suzuki, Y., and Kishita, K., Development of a gas injection/specimen heating holder for use with transmission electron microscope, *J. Electron Microsc.*, 2005, vol. 54, no. 6, pp. 497–503.

59. Sharma, R., Design and applications of environmental cell transmission electron microscope for in situ observations of gas-solid reactions, *Microsc. Microanal.*, 2001, vol. 7, no. 6, pp. 494–506.

60. Kawasaki, T., Ueda, K., Ichihashi, M., and Tanji, T., Improvement of windowed type environmental-cell transmission electron microscope for in situ observation of gas-solid interactions, *Rev. Sci. Instrum.*, 2009, vol. 80, no. 11, 113701. doi: 10.1063/1.3250862

61. Robertson, I.M. and Tetter, D., Controlled environment transmission electron microscopy, *Microsc. Res. Tech.*, 1998, vol. 42, no. 4, pp. 260–269.

62. Danilatos, G.D. and Postle, R., The environmental scanning electron microscope and its applications, *Scanning Electron Microsc.*, 1982, vol. 1, pp. 1–16.

63. Coillot, D., Podor, R., Méar, F.O., and Montagne, L., Characterization of self-healing glassy composites by high-temperature environmental scanning electron microscopy (HT-ESEM), *J. Electron Microsc.*, 2010, vol. 59, no. 5, pp. 359–366.

64. Sigee, D.C. and Gilpin, C., X-ray microanalysis with the environmental scanning electron microscope: Interpretation of data obtained under different atmospheric conditions, *Scanning Microsc., Suppl.*, 1994, vol. 8, pp. 219–227.

65. Hansen, T.W. and Wagner, J.B., Environmental transmission electron microscopy in an aberration-corrected environment, *Microsc. Microanal.*, 2012, vol. 18, no. 4, pp. 684–690.

66. Gauvin, R., Griffin, B., Nockolds, C., Phillips, M., and Joy, D.C., A method to measure the effective gas path length in the environmental or variable pressure scanning electron microscope, *Scanning*, 2002, vol. 24, no. 4, pp. 171–174.

67. Toth, M., Thiel, B.L., and Knowles, W.R., Gas cascade amplification in ultra-high-resolution environmental scanning electron microcopy, *Microsc. Microanal.*, 2010, vol. 16, no. 6, pp. 805–809.

68. Crozier, P.A. and Chenna, S., In situ analysis of gas composition by electron energy-loss spectroscopy for environmental transmission electron microscopy, *Ultramicroscopy*, 2011, vol. 111, no. 3, pp. 177–185.

69. Zou, Q. and Hatta, A., Electron field emission from carbon nanotubes in air for excitation of atmospheric pressure microplasma, *J. Nanosci. Nanotechnol.*, 2009, vol. 9, no. 2, pp. 924–928.

70. Kim, E.L., Chuprunova, S.E., and Portnov, V.N., A method for quality control of water-soluble single crystals by X-ray diffractometry under heating, *Ind. Lab. (Diagn. Mater.)*, 2000, vol. 66, no. 9, pp. 590–593.

71. Gukhman, A.A., Volynets, A.Z., Gavrilova, E.V., and Efremenko, G.N., Theory of psychrometry, *J. Eng. Phys.* (N. Y.), 1975, vol. 28, no. 4, pp. 499–504.

72. Nonami, H. and Boyer, J.S., Origin of growth-induced water potential: solute concentration is low in apoplast of enlarging tissues, *Plant Physiol.*, 1987, vol. 83, no. 3, pp. 596–601.

73. Shackel, K.A., Direct measurement of turgor and osmotic potential in individual epidermal cells: independent confirmation of leaf water potential as determined by in situ psychrometry, *Plant Physiol.*, 1987, vol. 83, no. 4, pp. 719–722.

74. Ham, R.G. and Puck, T.T., A regulated incubator controlling $CO_2$ concentration, humidity and temperature for use in animal cell culture, *Proc. Soc. Exp. Biol. Med.*, 1962, vol. 111, no. 1, pp. 67–71.

75. Liebes, L.F., Maher, V.M., Scherr, P., and McCormick, J.J., Automatic gas tank switching system for $CO_2$ incubators, *In vitro*, 1976, vol. 12, no. 3, pp. 265–268.

76. Ozawa, M., Nagai, T., Kaneko, H., Noguchi, J., Ohnuma, K., and Kikuchi, K., Successful pig embryonic development in vitro outside a $CO_2$ gas-regulated incubator: Effects of pH and osmolality, *Theriogenology*, 2006, vol. 65, no. 4, pp. 860–869.

77. Swif, J.A. and Brown A.C., An environmental cell for the examination of wet biological specimens at atmospheric pressure by transmission scanning electron microscopy, *J. Phys. E: Sci. Instrum.,* 1970, vol. 3, no. 11, pp. 924–926.

78. Parsons, D.F., Structure of wet specimens in electron microscopy, *Science,* 1974, vol. 186, no. 4162, pp. 407–414.

79. Donald, A.M., The use of environmental scanning electron microscopy for imaging wet and insulating materials, *Nat. Mater.,* 2003, vol. 2, no. 8, pp. 511–516.

80. Henderson, R. and McMullan, G., Problems in obtaining perfect images by single-particle electron cryomicroscopy of biological structures in amorphous ice, *Microscopy* (Oxford, U. K.), 2013, vol. 62, no. 1, pp. 43–50.

81. Nolin, F., Michel, J., Wortham, L., Tchelidze, P., Balossier, G., Banchet, V., Bobichon, H., Lalun, N., Terryn, C., and Ploton, D., Changes to cellular water and element content induced by nucleolar stress: Investigation by a cryo-correlative nano-imaging approach, *Cell. Mol. Life Sci.,* 2013, vol. 70, no. 13, pp. 2383–2394.

82. Krichevskii, E.S., Volchenko, A.G., Podgornyi, Yu.V., Proskuryakov, R.M., and Roskin, V.I., Thermovacuum aquametry—new method of measuring moistness, *Meas. Tech.,* 1976, vol. 19, no. 7, pp. 1042–1045.

83. Jung, Y.C. and Bhushan, B., Wetting behaviour during evaporation and condensation of water microdroplets on superhydrophobic patterned surfaces, *J. Microsc.* (Oxford, U. K.), 2008, vol. 229, no. 1, pp. 127–140.

84. De Mets, M. and Lagasse, A., Scanning electron microscopic investigation of gas chromatographic support material, *Chromatographia,* 1969, vol. 2, no. 9, pp. 401–403.

85. Sakodynsky, K. and Panina, L., Scanning electron microscopic investigations of gas chromatographic porous polymer sorbents, *Chromatographia,* 1971, vol. 4, no. 3, pp. 113–118.

86. Wong, C. and Yang, R.T., Scanning electron microscopy study of the kinetics of a gas-solid reaction, *Ind. Eng. Chem. Fundam.,* 1983, vol. 22, no. 4, pp. 380–384.

87. Kim, B., Lee, J.-g., Kim, E., Yun, S., Kim, K., and Kim, J.-Y., MFM and gas adsorption isotherm analysis of proton beam irradiated multi-walled carbon nanotubes, *Ultramicroscopy,* 2008, vol. 108, no. 10, pp. 1228–1232.

88. Sharma, R. and Weiss, K., Development of a TEM to study in situ structural and chemical changes at an atomic level during gas-solid interactions at elevated temperatures, *Microsc. Res. Tech.,* 1998, vol. 42, no. 4, pp. 270–280.

89. Nagoya University: Improvement of windowed type environmental-cell transmission electron microscope for in situ observation of gas-solid interactions, *Issues in Applied, Analytical, and Imaging Sciences Research,* Acton, Q.A., Ed., Atlanta, GA: ScholarlyEditions, 2011, 2011 ed.

90. Sayagués, M.J., Krumeich, F., and Hutchison, J.L., Solid-gas reactions of complex oxides inside an environmental high-resolution transmission electron microscope, *Micron,* 2001, vol. 32, no. 5, pp. 457–471.

91. Watson, J.H.L., Electron microscope observations of the morphology of several gases polymerized by charged-particle bombardment, *J. Phys. Colloid Chem.,* 1947, vol. 51, no. 3, pp. 654–661.

92. Barthlott, W., Ehler, N., and Schill R., Abtragung biologischer Oberflächen durch hochfrequenzaktivierten Sauerstoff für die Raster-Elektronenmikroskopie, *Mikroskopie,* 1976, vol. 32, nos. 1–2, pp. 35–44.

93. Ganczarczyk, A., Geller, M., and Lorke, A., $XeF_2$ gas-assisted focused-electron-beam-induced etching of GaAs with 30 nm resolution, *Nanotechnology,* 2011, vol. 22, no. 4, 045301. doi: 10.1088/0957-4484/22/4/045301

94. Wu, H.M., Stern, L.A., Chen, J.H., Huth, M., Schwalb, C.H., Winhold, M., Porrati, F., Gonzalez, C.M., Timilsina, R., and Rack, P.D., Synthesis of nanowires via helium and neon focused ion beam induced deposition with the gas field ion microscope, *Nanotechnology,* 2013, vol. 24, no. 17, 175302. doi: 10.1088/0957-4484/24/17/175302

95. Adamiak, B. and Mathieu, C., The reduction of the beam gas interactions in the variable pressure scanning electron microscope with the use of helium gas, *Scanning,* 2000, vol. 22, no. 3, pp. 178–181.

96. Mansour, O., Kadoun, A., Khouchaf, L., and Mathieu, C., Monte Carlo simulation of the electron beam scattering under water vapor environment at low energy, *Vacuum,* 2013, vol. 87, pp. 11–15.

97. Chee, A.K.W., Broom, R.F., Humphreys, C.J., and Bosch, E.G.T., A quantitative model for doping contrast in the scanning electron microscope using calculated potential distributions and Monte Carlo simulations, *J. Appl. Phys.* (Melville, NY, U. S.), 2011, vol. 109, no. 1, 013109. doi: 10.1063/1.3524186

98. Hoburg, A., Keshlaf, S., Schmidt, T., Smith, M., Gohs, U., Perka, C., Pruss, A., and Scheffler, S., Fractionation of high-dose electron beam irradiation of BPTB grafts provides significantly improved viscoelastic and structural properties compared to standard gamma irradiation, *Knee Surg. Sports Traumatol. Arthrosc.,* 2011, vol. 19, no. 11, pp. 1955–1961.

99. Rubino, S., Akhtar, S., Melin, P., Searle, A., Spellward, P., and Leifer, K., A site-specific focused-ion-beam lift-out method for cryo transmission electron microscopy, *J. Struct. Biol.,* 2012, vol. 180, no. 3, pp. 572–576.

100. Khouchaf, L., The surface skirt in gaseous scanning electron microscope (GSEM), *Microsc. Res.,* 2013, vol. 1, no. 3, pp. 29–32.

101. Khouchaf, L. and Verstraete, J., Electron scattering by gas in the environmental scanning electron microscope (ESEM): Effects on the image quality and on the X-ray microanalysis, *J. Phys. IV,* 2004, vol. 118, pp. 237–243.

102. Shah, J.S. and Beckett, A., A preliminary evaluation of moist environment ambient temperature scanning electron microscopy, *Micron,* 1979, vol. 10, no. 1, pp. 13–23.

103. Fletcher, A.L., Thiel, B.L., and Donald, A.M., Signal components in the environmental scanning electron

microscope, *J. Microsc.* (Oxford, U. K.), 1999, vol. 196, no. 1, pp. 26−34.

104. Erlykin, A., Sloan, T., and Wolfendale, A., Cosmic rays, climate and the origin of life, CERN Courier. http://cerncourier.com/cws/article/cern/41723.

105. Erlykin, A.D. and Wolfendale, A.W., Long term time variability of cosmic rays and possible relevance to the development of life on earth, *Surv. Geophys.,* 2010, vol. 31, no. 4, pp. 383−398.

106. Stribeck, N., *X-ray Scattering of Soft Matter*, Berlin−Heidelberg: Springer-Verlag, 2007.

107. Mulkidjanian, A.Y. and Galperin, M.Y., On the origin of life in the zinc world. 2. Validation of the hypothesis on the photosynthesizing zinc sulfide edifices as cradles of life on Earth, *Biol. Direct,* 2009, vol. 4, no. 27. doi: 10.1186/1745-6150-4-27

108. Podkletnov, N.E. and Markhinin, E.K., New data on abiogenic synthesis of prebiological compounds in volcanic processes, *Origins Life,* 1981, vol. 11, no. 4, pp. 303−315.

109. Balucani, N., Nitrogen fixation by photochemistry in the atmosphere of titan and implications for prebiotic chemistry, *The Early Evolution of the Atmospheres of Terrestrial Planets,* Astrophys. Space Sci. Proc., New York: Springer, 2013, vol. 35, pp. 155−164.

110. Abel, D.L., Moving "far from equilibrium" in a prebiotic environment: The role of Maxwell's Demon in life origin, *Genesis—In The Beginning: Precursors of Life, Chemical Models and Early Biological Evolution,* Cell. Origin, Life Extreme Habitats Astrobiol., New York: Springer, 2012, vol. 22, pp. 219−236.

111. Kobayashi, K., Tsuchiya, M., Oshima, T., and Yanagawa, H., Abiotic synthesis of amino acids and imidazole by proton irradiation of simulated primitive earth atmosphere, *Origins Life Evol. Biospheres,* 1990, vol. 20, no. 2, pp. 99−109.

112. Kobayashi, K., Kaneko, T., Tsuchiya, M., Saito, T., Yamamoto, K., Koike, J., and Oshima, T., Formation of bioorganic compounds in planetary atmospheres by cosmic radiation, *Adv. Space Res.,* 1995, vol. 15, no. 3, pp. 127−130.

113. Kobayashi, K., Kaneko, T., Saito, T., and Oshima, T., Amino acid formation in gas mixtures by high energy particle irradiation, *Origins Life Evol. Biospheres,* 1998, vol. 28, no. 2, pp. 155−165.

114. Simakov, M.B., Kuzicheva, E.A., and Gontareva, N.B., Abiogenic synthesis of oligopeptides in the open space, *Paleontol. J.,* 2013, vol. 47, no. 9, pp. 1097−1103.

115. Martell, E.A., Radionuclide-induced evolution of DNA and the origin of life, *J. Mol. Evol.,* 1992, vol. 35, no. 4, pp. 346−355.

116. Tamulis, A. and Grigalavicius, M., Quantum entanglement in photoactive prebiotic systems, *Syst. Synth. Biol.,* 2014, vol. 8, no. 2, pp. 117−140.

117. Tamulis, A., Grigalavicius, M., and Baltrusaitis, J., Phenomenon of quantum entanglement in a system composed of two minimal protocells, *Origins Life Evol. Biospheres,* 2013, vol. 43, no. 1, pp. 49−66.

118. Bakr, W.S., Gillen, J.I., Peng, A., Fölling, S., and Greiner, M., A quantum gas microscope for detecting single atoms in a Hubbard-regime optical lattice, *Nature,* 2009, vol. 462, no. 7269, pp. 74−77.

119. Gericke, T., Würtz, P, Reitz, D., Langen, T., and Ott, H., High-resolution scanning electron microscopy of an ultracold quantum gas, *Nat. Phys.,* 2008, vol. 4, no. 12, pp. 949−953.

*Translated by E. Boltukhina*